# Effect of the alkyl chain length of the alcohols on the nematic uniaxial-to-biaxial phase transitions in the potassium laurate/alcohol/K$_2$SO$_4$/water lyotropic mixture

*Erol Akpinar[1], Dennys Reis[2] and Antonio Martins Figueiredo Neto[2]\**

[1] Abant Izzet Baysal University, Arts and Sciences Faculty, Department of Chemistry, 14280, Bolu, TURKEY

[2] Universidade de São Paulo, Instituto de Física, caixa postal 66318, 05314-970, São Paulo, SP, BRASIL

\* Author for correspondence e-mail: afigueiredo@if.usp.br

**RECEIVED DATE (to be automatically inserted after your manuscript is accepted if required according to the journal that you are submitting your paper to)**

ABSTRACT. Lyotropic liquid crystalline quaternary mixtures of potassium laurate (KL), potassium sulphate ($K_2SO_4$)/alcohol (n-OH)/water, with the alcohols having different number of carbon atoms in the alkyl chain ($n$), from 1-octanol to 1-hexadecanol, were investigated by optical techniques (optical microscopy and laser conoscopy). The biaxial nematic phase domain is present in a window of values of $n = n_{KL} \pm 2$, where $n_{KL} = 11$ is the number of carbon atoms in the alkyl chain of KL. The biaxial phase domain got smaller and the uniaxial-to-biaxial phase transition temperatures shifted to the relatively higher temperatures on going from 1-nonanol to 1-tridecanol. Moreover, these new mixtures present high values of the birefringences comparing to other lyotropic mixtures. This result is expected to be related to the micellar shape anisotropy. Our results are interpreted assuming that alcohol molecules tend to segregate in the micelles in a way that depends on the relative value of $n$ with respect to $n_{KL}$. The larger the value of $n$, the more alcohol molecules tend to be located in the curved parts of the micelle, favoring the uniaxial nematic calamitic phase with respect to the biaxial and uniaxial discotic nematic phases.

INTRODUCTION

After the experimental realization of the biaxial nematic phase ($N_B$) in the lyotropic mixture of potassium laurate (KL)/1-decanol (DeOH)/water by Yu and Saupe in 1980 [1], some research groups reported on new mixtures presenting the $N_B$ phase during the last three decades [2-9]. These studies were useful to the understanding of the physical-chemical characteristics of this remarkable phase, about its chemical stability and the symmetry of the micelles in these mixtures. For instance, some authors proposed that the biaxial phase should be made of the co-existence of two uniaxial nematic discotic $N_D$ and calamitic $N_C$ phases. From the nanoscopic point of view, this idea leads to the so-called model of "mixtures of cylindrical-like and disc-like micelles, MCD" [10]. This type of model had many problems [11-13], and a different approach was proposed by Neto *et al.* [14] and Galerne *et al.* [15].

They proposed the intrinsically biaxial micelles (IBM) model, which is based on the different orientational fluctuations of the same type of micelles in the three nematic phases.

One of the most easy and reliable experimental method to investigate the uniaxial-to-biaxial phase transitions is the laser conoscopy [16,17]. This technique is particularly useful in the case of materials with small birefringence ($\sim 10^{-3}$), where thick samples are needed to allow a measurement of the birefringences with accuracy.

Macroscopically the $N_B$ phase has orthorhombic symmetry and, consequently, three different indices of refraction, along the three orthogonal symmetry axes, each one of these axes of order two [18]. As a consequence, two birefringences are present, defined as the differences between two of those principal indices of refraction. The nematic phases are characterized by an order-parameter that is a second-rank, symmetric and traceless tensor. Practically, the optical dielectric tensor may be chosen as the order-parameter and it can be calculated from the measurements of the birefringences [17]. Moreover, the temperature dependences of the birefringences are very useful for the determination of the phase transition temperatures between the uniaxial-to-biaxial nematics, and also to build up phase diagrams.

On comparing the phase diagrams of lyotropic mixtures published in the literature up-to-date, they gave, in general, similar biaxial phase properties in terms of the temperature range of the biaxial phase and maximum value of the nematic birefringence ($\sim 2 \times 10^{-3}$) [19]. Assuming the IBM model it is expected that the higher the birefringence value, the more anisometric the micelles should be. An interesting result encountered so far is that in all the lyotropic mixtures presenting the $N_B$ phase there are always a surfactant and a co-surfactant [18]. Mixtures with only one surfactant were shown to present only one of the uniaxial nematic phases, the $N_C$ or the $N_D$ phase. The alcohol is commonly employed [1,3] as a co-surfactant, however, mixtures with soap and a detergent [5] also presented the $N_B$ phase.

Let us now concentrate our attention in the alcohol molecules present in lyotropic mixtures. Consider a micellar systems composed by an amphiphilic compound, water and salt. In this system three different regions may be identified [20]: the first one is the aqueous region (intermicellar region) that contains dissolved counterions of the surfactants, a small amount of free amphiphiles and the dissociated salt ions; the second region (interfacial region or palisade layer [21]) consists of the headgroups of the surfactants, some amount of counterions and water; the third region (hydrophobic core) is composed of the surfactant alkyl chains. When alcohol molecules are added to this mixture they may be located in these regions, depending on their characteristics. Alcohol molecules of short chain, such as ethanol, are completely miscible with water and it is mostly found in the intermicellar and in the palisade regions. However, they cannot efficiently penetrate into the micellar core. Then, the mole fraction of ethanol incorporated in the micelle should be smaller than that in the intermicellar. The mole fraction of alcohol molecules incorporated in the micelles ($X_A$) may be different from the mole fraction in the intermicellar region ($X_B$). The ratio between these two quantities is known as the partition coefficient of the alcohol, $K_p = X_A / X_B$ [22,23]. The bigger the number of carbon atoms in the alkyl chain of the alcohol molecule, the bigger $K_p$ [22-26]. Since short chain alcohols have higher solubility in the water than long chain alcohols, short chain alcohols have smaller $K_p$ values than longer chain alcohols. In other words, while the short chain alcohols preferentially exist in the intermicellar region, long chain alcohols are mainly incorporated to the micelles [21,27]. For example, the $K_p$ value of hexanol is 758, approximately 18 times that of butanol ($K_p = 42.2$) in aqueous micellar solutions of lithium dodecylsulfate [28]. In lipid membrane/buffer systems, the partition coefficients (defined here as the ratio of molar concentration of alcohol in the lipid bilayer and in the buffer - or water layer) of decanol, undecanol, dodecanol, tridecanol, tetradecanol and pentadecanol are $3.9 \times 10^3$; $1.7 \times 10^4$; $6.9 \times 10^4$; $3.1 \times 10^5$; $1.4 \times 10^6$ and $5.6 \times 10^6$, respectively [29]. This means that, in terms of the molar concentrations or number of alcohol molecules in the bilayer (or in the hydrophophic core), pentadecanol incorporates much more molecules than decanol.

To the best of our knowledge there is not in the literature a systematic investigation of the effect number of carbons in the alkyl chain of the alcohol molecules ($n$) in the topology of lyotropic mixtures presenting the biaxial nematic phase. In this study we investigate the phase sequence of ten new lyotropic mixtures of KL/K$_2$SO$_4$/1-nOH/water, with $n = 6$ and $8 \leq n \leq 16$, determining the location of the N$_B$ phase. Optical microscopy and laser conoscopy experimental techniques are used to characterize the phases.

EXPERIMENTAL

The sample preparation, in particular its homogenization procedure, was discussed elsewhere [18]. This aspect is extremely important to get reproducible results. To improve the orientation of the samples in the magnetic field, a small amount (~1 μL ferrofluid per 1 g of mixture) of water-base ferrofluid was added to the mixtures [18]. For the observation of the textures under a light polarized optical microscope, each sample was transferred into a 0.2 mm thick flat-glass microslide.

Besides the texture characterization, the nematic sample's birefringences were measured by using laser conoscopy [17]. The laboratory frame axes were defined as follows: the horizontal plane is defined by the two orthogonal axes 1 and 2; the magnetic field is aligned along the 1 axis; axis 3 is vertical and parallel to the laser beam propagation direction. Two main optical birefringences, $\Delta n = n_2 - n_1$ and $\delta n = n_3 - n_2$, were measured as a function of temperature. Mixtures were put in the sample cell made of two circular optical glasses and a glass o-ring, leaving a liquid crystalline film 1.0 mm thick. Sample's birefringences were measured as a function of temperature (T). The experimental setup has a HeNe laser ($\lambda = 632.8\,nm$) and a Neocera LTC-21 temperature control, with accuracy of 0.001°C. A Walker Sci. electromagnet provides the static magnetic field $\left(\left|\vec{H}\right| = 2.04\,kG\right)$ used to orient the sample.

To obtain well-aligned N$_D$ phase, the samples were kept at a fixed temperature for a period of time of about 30-60 min. After that period the temperature was slowly varied step by step. At each new temperature the sample was left at rest for at least 10 min in order to achieve the thermal equilibrium. From time to time, in the $N_D$ ($N_B$) phase, the sample was turned of an angle of about $\pm \pi/2$ ($\pm \pi/6$) around axis 3 to improve the sample orientation.

Mixtures are named according to the number of carbon atoms in the alkyl chain, e.g., $n_{10}$ corresponds to the KL/K$_2$SO$_4$/1-decanol/water mixture, $n_{16}$ corresponds to the KL/K$_2$SO$_4$/1-hexadecanol/water, and so on. Table 1 presents the molar fractions of the different mixture compounds; the nematic phases identified and the temperature range of the biaxial nematic phase, when it is present.

**Table 1:** Molar fractions ($X$) of the components of the lyotropic mixtures investigated. $\Delta T_B$ is the temperature range of the biaxial nematic domain.

| Mixture | Alcohol | $X_{KL}$ | $X_{K_2SO_4}$ | $X_{alcohol}$ | $X_{H_2O}$ | Nematic Phase type | $\Delta T_{N_B}/°C$ |
|---|---|---|---|---|---|---|---|
| $n_6$ | HeOH | 0.0383 | 0.0060 | 0.0114 | 0.9443 | --- | --- |
| $n_8$ | OcOH | 0.0382 | 0.0060 | 0.0114 | 0.9443 | $N_D$ | --- |
| $n_9$ | NonOH | 0.0383 | 0.0060 | 0.0114 | 0.9443 | $N_B, N_D$ | > 6 |
| $n_{10}$ | DeOH | 0.0383 | 0.0060 | 0.0114 | 0.9443 | $N_C, N_B, N_D$ | ~7.80 |
| $n_{11}$ | UnDeOH | 0.0383 | 0.0060 | 0.0114 | 0.9443 | $N_C, N_B, N_D$ | ~5.55 |
| $n_{12}$ | DDeOH | 0.0383 | 0.0060 | 0.0114 | 0.9443 | $N_C, N_B, N_D$ | ~3.30 |
| $n_{13}$ | TDeOH | 0.0383 | 0.0060 | 0.0114 | 0.9443 | $N_C, N_B, N_D$ | ~1.70 |
| $n_{14}$ | TeDeOH | 0.0382 | 0.0060 | 0.0114 | 0.9443 | $N_C$ | --- |
| $n_{15}$ | PDeOH | 0.0382 | 0.0060 | 0.0114 | 0.9443 | $N_C$ | --- |
| $n_{16}$ | HDeOH | 0.0383 | 0.0060 | 0.0114 | 0.9443 | $N_C$ | --- |

# RESULT AND DISCUSSION

The textures of the mixtures were identified under polarizing light microscope. In general, the textures observed were similar to those from typical nematic lyotropic mesophases [18]. The different nematic phases and phase-transition temperatures were obtained from the birefringences measurements and texture identification.

Since the mixtures from the $n_9$ until the $n_{13}$ show the $N_D$ phase at higher temperatures, the well-oriented $N_D$ phases were obtained by applying the static magnetic field on the sample, and then, the optical birefringences of each nematic phase as a function of temperature were measured, starting from the $N_D$ phase, upon cooling. The experimental results of $\Delta n$ and $\delta n$ as a function of T, from these mixtures, keeping constant the molar fractions of all constituents in all mixtures (see Table 1), are shown in Figure 1. It is interesting to note that, going from $n_9$ to $n_{13}$, the uniaxial-to-biaxial phase transitions shift to higher temperatures and the biaxial nematic domain gets smaller.

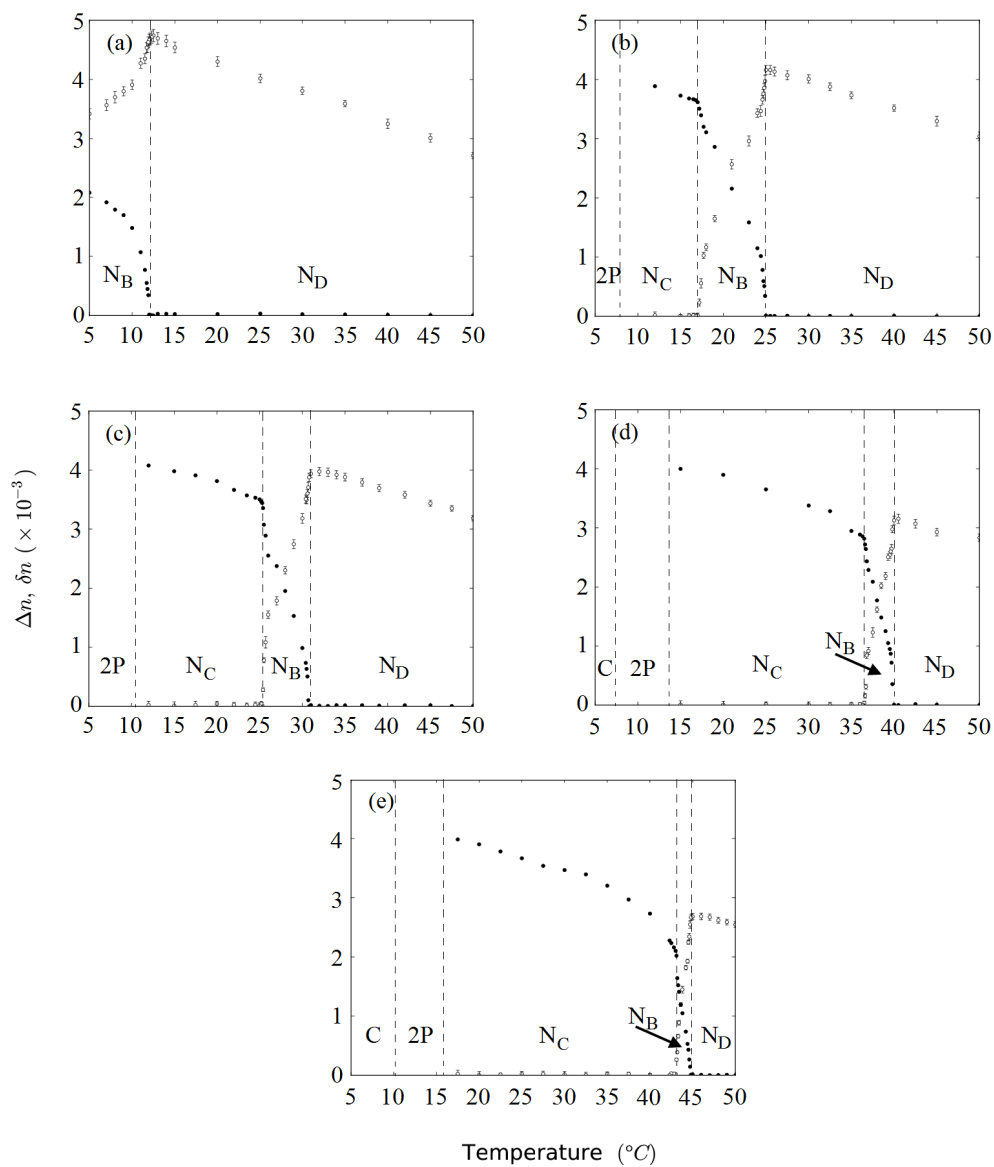

**Figure 1.** Birefringences of the mixtures of (a) $n_9$, (b) $n_{10}$, (c) $n_{11}$, (d) $n_{12}$, and (e) $n_{13}$, as a function of the temperature. $\Delta n = n_2 - n_1$ (●) and $\delta n = n_3 - n_2$ (○). The labels 2P and C stand for a two-phase and crystalline regions, respectively.

The phase diagram depicted in Fig. 2 was constructed by plotting the different phases identified in the surface temperature (T) *versus* the number of the carbon atoms $(n)$ in the alkyl chain of the alcohols. Special care must be taken in the conclusions drawn from the plot shown in Fig. 2 because the temperature assumes continuous values (intensive thermodinamical variable), and $n$ assumes only a discrete set of positive values. The data were obtained from both the laser conoscopy and optical polarizing microscopy. The existence of the three nematic phases is restricted to the range of values of $n$, from $n = 9$ to $n = 13$. The mixture with 1-hexanol was also prepared, however, this mixture was highly viscous and its homogeneity was hardly achieved. The visual inspection seems to indicate the presence of a gel phase in all the temperature range investigated.

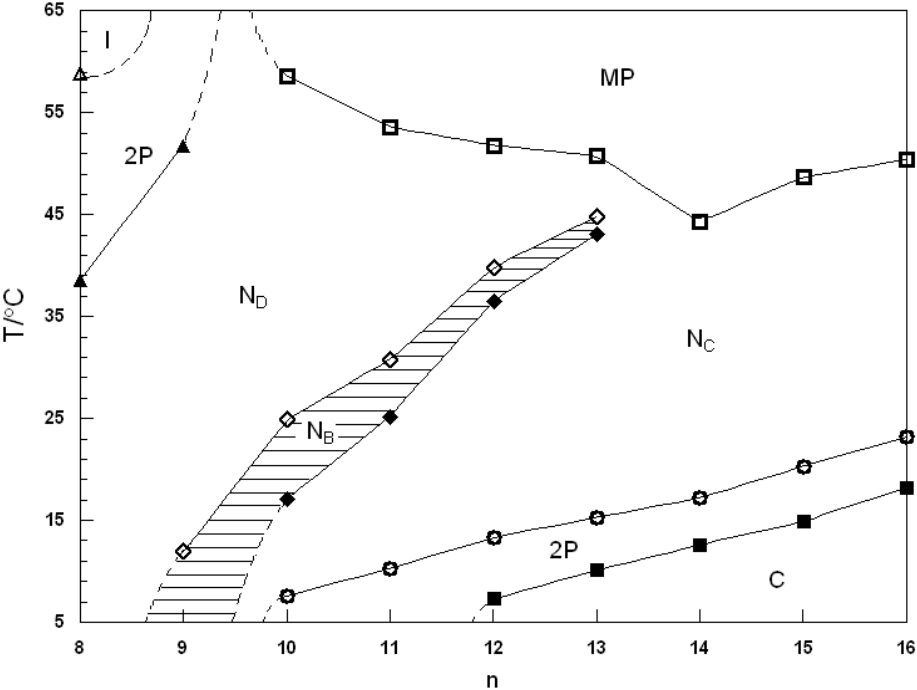

**Figure 2.** Phase diagram of the lyotropic mixtures: temperature *versus* the number of carbon atoms in the alkyl chain of the alcohols. I, 2P, MP and C stand for an isotropic phase, two-phase region, multi-phase region and crystalline phase, respectively. Solid and dashed lines are only guides for the eyes. The hatched region depicts the biaxial nematic phase domain.

The optical birefringences in these new quaternary lyotropic mixtures are higher when comparing with those from the usual ternary KL/DeOH/water mixture. In particular, in the $N_D$ phase of our quaternary mixtures with nonanol, decanol and undecanol, birefringence reaches values of bigger than $4 \times 10^{-3}$, almost two times the maximum value encountered in the KL/DeOH/water system [19]. This high birefringence values are not due to the ferrofluid doping. We check this by measuring the birefringences of (doped) sample with $n=8$ in the isotropic phase: within our experimental accuracy $\Delta n = \delta n = 0$. Going from the 1-decanol to 1-dodecanol, the $N_B$ and $N_D$ domains got smaller and the $N_C$ domain increases (Figure 3).

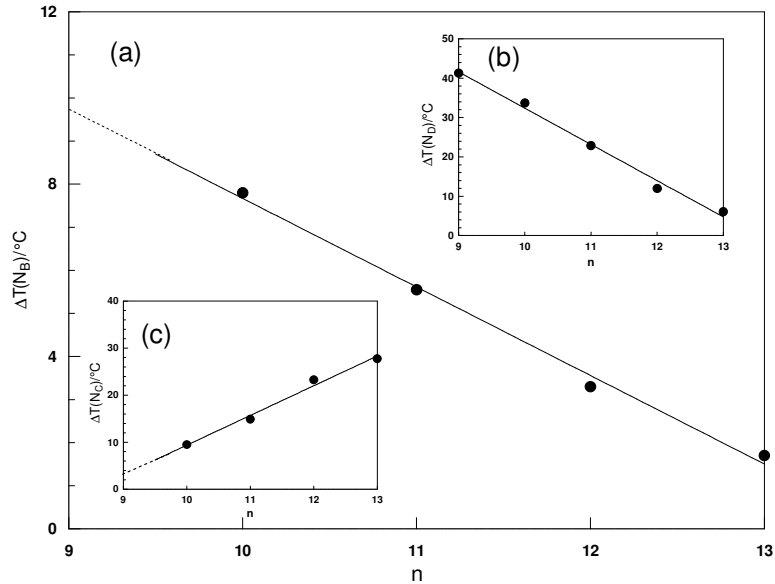

**Figure 3.** Temperature intervals of the $N_B$ (a), $N_D$ (b) and $N_C$ (c) nematic phases of the different quaternary mixtures, where n represents the number of carbons on the alkyl chains of the alcohols. Solid lines are linear fits.

The nematic phases of the lyotropic liquid crystals are characterized by a second-rank, traceless, symmetric-tensor order parameter [30,31]. The optical dielectric tensor, $\vec{\varepsilon}$, which has the same symmetry of the phases, may be chosen as the order parameter [32]. The symmetric invariants of the tensor order parameter, $\sigma_1$, $\sigma_2$ and $\sigma_3$ are calculated from the birefringences measurements [17]. The first invariant $\sigma_1$ is simply the trace of the tensor, which, in this case, is equal to zero. In the uniaxial phases the invariant $\sigma_3$ is related to $\sigma_2$ by $\sigma_3 = \pm \sigma_2^{3/2}$ [17]. The signs + (−) refer to the $N_D$ ($N_C$) phase. In the biaxial phase $-\sigma_2^{3/2} < \sigma_3 < \sigma_2^{3/2}$. Figures 4 and 5 show the dependences of $\sigma_2$ and $\sigma_3$ on temperature of the mixtures presenting the $N_B$ phase, respectively. The mean-field theory predicts that, near the uniaxial-to-biaxial phase transitions, $\sigma_2$ and $\sigma_3$ depend linearly on T. This behavior is observed on Figs. 4 and 5. Figure 6 shows the $\sigma_3$ values plotted against $\sigma_2$, and the solid lines represent the $\sigma_3$ versus $\sigma_2$ theoretical behavior for the uniaxial nematic phases. The invariants of the order parameter also reach high values.

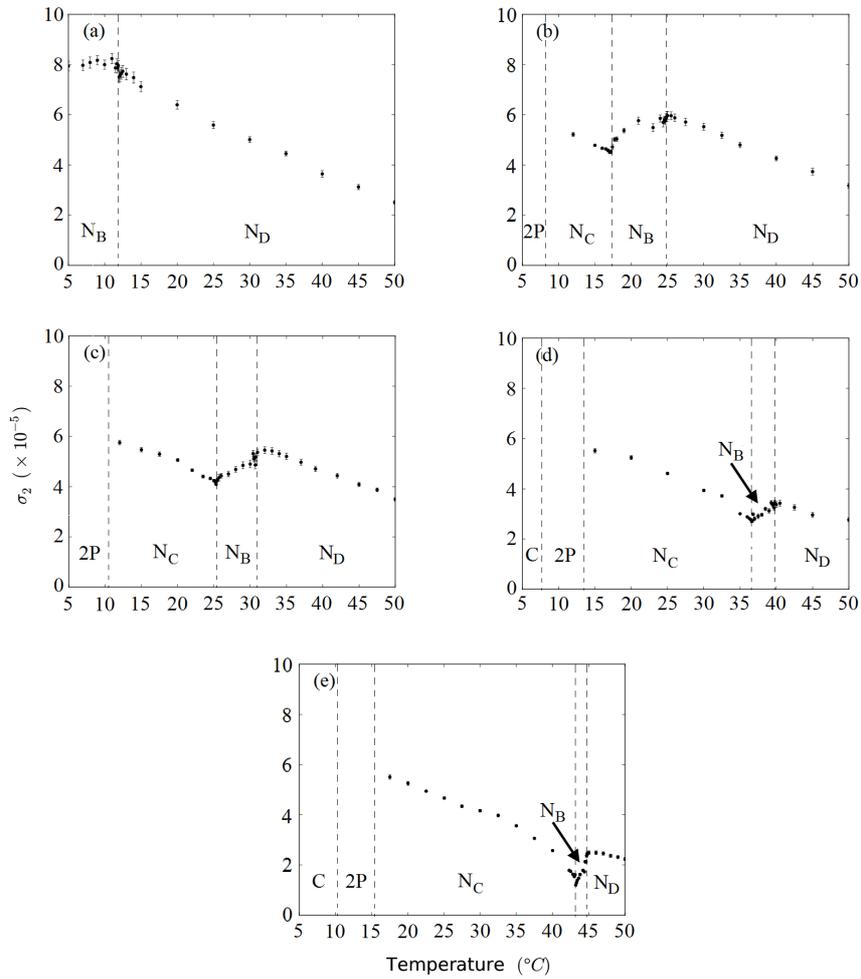

**Figure 4.** Order-parameter invariant $\sigma_2$ versus T: (a) 1-nonanol, (b) 1-decanol, (c) 1-undecanol, (d) 1-dodecanol and (e) 1-tridecanol mixtures. 2P and C are the two-phase and crystalline phase regions, respectively.

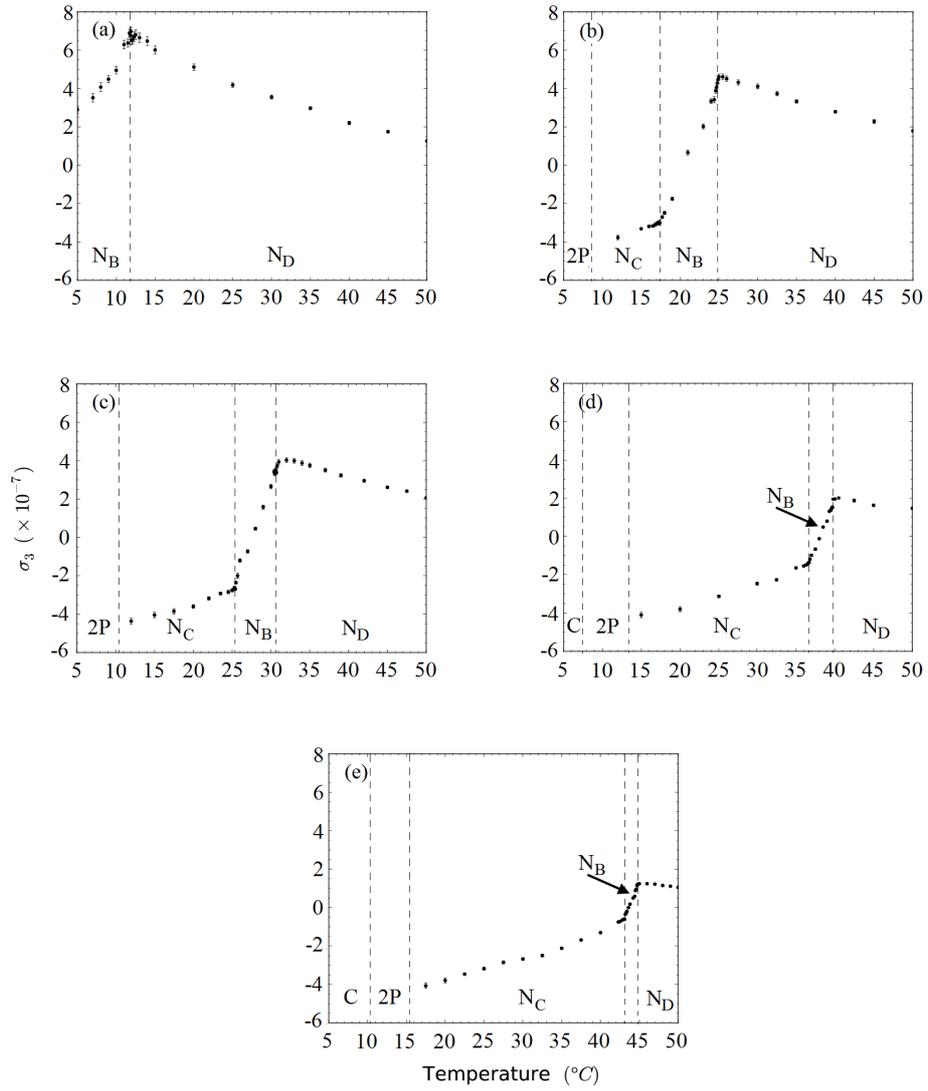

**Figure 5.** Order-parameter invariant σ₃ *versus* T: (a) 1-nonanol, (b) 1-decanol, (c) 1-undecanol, (d) 1-dodecanol and (e) 1-tridecanol mixtures. 2P and C are the two-phase and crystalline phase regions, respectively.

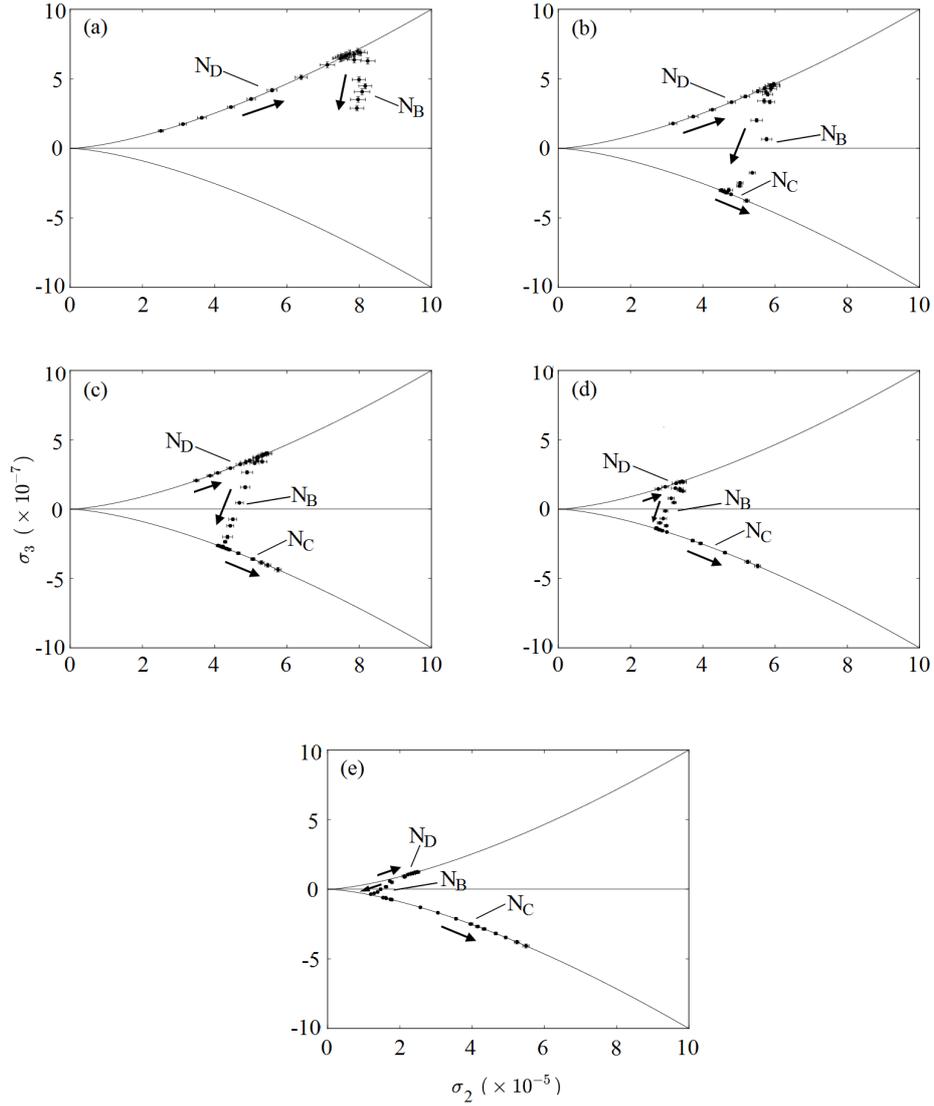

**Figure 6.** Dependence of the invariant $\sigma_3$ with respect to $\sigma_2$ for the mixtures (a) 1-nonanol, (b) 1-decanol, (c) 1-undecanol, (d) 1-dodecanol and (e) 1-tridecanol, at the working temperature range. Solid lines represent the mean-field $\sigma_3$ versus $\sigma_2$ theoretical behavior for the uniaxial nematic phases. The arrows indicate the cooling process employed to measure the birefringences, starting from the $N_D$ phase.

In our case, all the alcohols used in this study present high partition coefficients $K_p$ [29]. Since they are completely immiscible in water, we have to discuss the locations of their head groups, -OH, in the palisade layer and the hydrophobic tails in the micelle core. In the framework of the IBM model [15,18], the higher the birefringences of the nematic phase, the larger the shape anisotropy of the micelles. Therefore, our new quaternary mixtures are expected to show more anisometric micelles when compared with those from the usual KL/DeOH/water system, since the maximum values of the birefringences measured in this ternary mixture lies between $3 \times 10^{-3}$ and $4 \times 10^{-3}$ in the nematic temperature range. Let us sketch a micelle as an object of orthorhombic symmetry, as in Fig. 7, with typical dimensions A', B' and C'. The axes of the coordinate system fixed in the micelles are $\alpha$, $\beta$ and $\gamma$, and the laboratory frame axes are 1, 2 and 3.

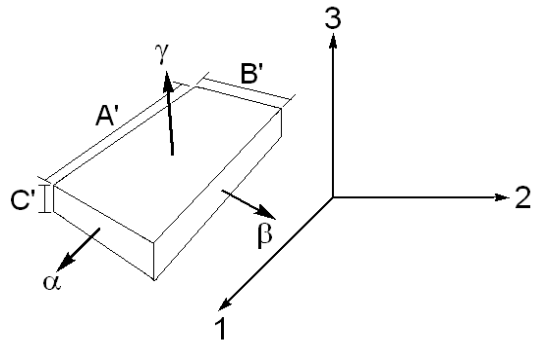

**Figure 7.** Sketch of the orthorhombic micelle, in the framework of the IBM model. The detergent amphiphilic bilayer is represented by C'.

According to neutron contrast experiments with the lyotropic mixture KL/DeOH/water, alcohol molecules concentrate, mainly, in the flattest surface of the micelle, rather than in its curved borders [33]. This result indicates that there is segregation on the partitions of the surfactant and alcohol molecules in the less curved and rims of the micelles. From the experimental point of view, the increase of the alcohol and water relative concentration in the mixture was shown to enlarge the $N_D$ phase domain [2]. This result may be interpreted as an increase of the dimensions of the micelles in the plane $\alpha - \beta$. This geometry favors the orientational fluctuation that degenerates axis 3, stabilizing the $N_D$ phase. In our present experiment the mole-fractions of the different compounds are constant in all the mixtures investigated. So, the effect of changing the different nematic phase-domains sizes is due to the relative difference between $n$ (number of carbon atoms in the alkyl chain of the different alcohols) and $n_{KL}$ (=11, number of carbon atoms in the alkyl chain of the detergent, except that placed at the polar head [34]). The $N_B$ phase appears in a window of $n = n_{KL} \pm 2$. When $n < 11$ our results show the prevalence of the $N_D$, with respect to other nematic phases. We expect that, in this case, alcohol molecules accumulates preferentially in the largest micellar surfaces, and the micellar dimensions are such that $A' \sim B'$. However, it must be pointed out that the dimensions A' and B' must be different to give rise to the $N_B$ phase, observed in these mixtures. This particular shape anisometry favors the orientational fluctuations that degenerates the 3-axis, giving rise to the $N_D$ phase. When $n > 11$, alcohol molecules may accommodate themselves also in the curved parts of the micelle in a larger number. By affinity, it is expected that these molecules tend to concentrate in a particular place of the micelle, increasing the shape anisotropy in the $\alpha - \beta$ plane. When $n > 11$, the presence of long chain alcohol molecules in the more flat part of the micelle (with bilayer of $C' = 2.6$ nm [15]) becomes more difficult, and the alcohol molecules tend to concentrate in the curved parts of the micelle, increasing still more the shape anisotropy in the $\alpha - \beta$ plane. This behavior favors the increase of one of the micellar dimensions, let us say A' with respect to B' in our sketch of Fig. 7, and the orientational fluctuation which is favored is the one which degenerates the 1-axis, i.e., privileging the $N_C$ phase. This accumulation of alcohol molecules in the more curved part of the micelle may be interpreted as a nanosegregation due to the affinity between identical long-chain molecules, giving rise to an increase of one of the micellar dimension (A') with respect to the other (B'). At this point we define a phenomenological partition coefficient (similar to that discussed in the Introduction section) that takes into account the process of alcohol segregation in the micelle employed here to interpret our experimental results. Let $K_p^m$ be the ratio between the mole fraction of alcohol molecules in the more curved surfaces, and that of the more flat surfaces of the micelle. The higher the $K_p^m$ value, the more alcohol molecules accumulate in the curved parts of the micelle. Our results seems to indicate that the increase of the carbons in the alcohol alkyl chain leads to an increase of $K_p^m$, since increasing $n$ the $N_C$ phase domain increases at the expense of the $N_D$ and $N_B$ phase domains.

CONCLUSIONS

We investigate the phase diagram of the quaternary mixture of KL/K$_2$SO$_4$/alcohol/water as a function of the temperature and number of carbon atoms in the alcohol alkyl chain ($n$), where the three nematic phases are present. We show that the biaxial nematic phase domain exists in a window of values of $n$ varying around the number of carbon atoms of the KL chain, with $n = n_{KL} \pm 2$. The higher the value of $n$ the larger the $N_C$ phase domain, when comparing to the $N_D$ and $N_B$ phase domains. Assuming that micelles have an orthorhombic symmetry (model of intrinsically biaxial micelles in the nematic phases), our results suggest that the alcohol molecules segregate in different ways, depending on the value of $n$ with respect to $n_{KL}$: for $n < n_{KL}$ there is a tendency of the alcohol molecules accumulate more in the flattest surface of the micelles, favoring the $N_D$ phase; for $n > n_{KL}$ alcohol molecules tend accommodate preferentially in the curved surfaces of the micelle, favoring the $N_C$ phase. This segregation of the alcohol molecules seems to maintain the orthorhombic symmetry of the micelles, probably due to the fact that similar molecules tend to remain together.

ACKNOWLEDGMENT. We thank the Scientific and Technological Research Council of Turkey (TÜBİTAK), the National Institute of Science and Technology on Complex Fluids (INCT-FCx), CNPq and FAPESP from Brazil for supporting this study.